\documentclass[useAMS,usenatbib,preprint]{mn2e}
\usepackage{graphicx}
\usepackage{pstricks}
\usepackage{psfrag}
\usepackage{wrapfig}
\usepackage{xspace}

\newcommand{\etal}{et al. }

\newcommand{\apj}{    {\it Astrophys. J.}}

\newcommand{\mnras}{  {\it Mon. Not. Roy. Astron. Soc.}}

\newcommand{\solphys}{{\it Solar Phys.}}
\newcommand{\sovast}{ {\it Sov. Astronom.}}

\title[A magneto-seismic study of the active region AR10720]{Helioseismic analysis of the solar flare-induced sunquake of 2005 January 15. II: A magneto-seismic study}
\author[J. C. Mart\'inez-Oliveros, A.-C. Donea, P.S. Cally, H. Moradi]{J. C. Mart\'inez-Oliveros\thanks{E-mail:
 Juan.Oliveros@sci.monash.edu.au}, A.-C. Donea, P.S. Cally, H. Moradi\\
Centre for Stellar and Planetary Astrophysics, School of Mathematical Sciences, Monash University, Victoria 3800, Australia}

\begin{document}

\date{Accepted 1988 December 15. Received 1988 December 14; in original form 1988 October 11}

\pagerange{\pageref{firstpage}--\pageref{lastpage}} \pubyear{2002}

\maketitle

\label{firstpage}

%
%
%
%
\begin{abstract}
On 2005 January 15, the active region AR10720
produced an X1.2 solar flare that induced high levels of seismicity into
the photospheric layers. The seismic source was detected using
helioseismic holography and analysed in detail in Paper~I. Egression power maps at 6 mHz with a 2 mHz
bandwidth revealed a compact acoustic source strongly correlated with
the footpoints of the  coronal loop that hosted the flare. We present
a magneto-siesmic study of this active region in order to understand, for the
first time, the magnetic topological structure of a coronal
field that hosts an acoustically active solar flare. The accompanying analysis attempts to answer questions such as: Can the magnetic field act as a barrier and prevent seismic waves from spreading away from the 
focus of the sunquake? And, what is the most {\em efficient} magnetic structure that would facilitate the development of a strong seismic source in the photosphere?
 
\end{abstract}



\section{Introduction}

Our understanding of the acoustics of solar flares has been greatly
improved in recent years through a combination of observational and
computational techniques. It was \citet{w1972} who first suggested that solar flares
could release acoustic noise into the solar interior. \citet{kz1995}
simulated this phenomenon for the first time, and soon after, \citet{kz1998}
discovered the first seismic event, in the form of ripples, propagating away from
the flare of 1996 July 9.  With the advancement of
local helioseismic techniques such as helioseismic holography
\citep{lb2000}, we have now detected numerous seismic sources of
varying size and intensity produced by M- and X-class flares
\citep{dbl1999,dl2005,donea2006,moradi07,betal2007,Martinez-Oliveros2007}. Extended
work on this field has also been continued  by \citet{kz2006, zharkova2007,
Martinez-Oliveros2008}.

During the impulsive phase of a flare, the coronal magnetic energy is
transferred down into the photosphere and further into the solar
interior. This energy is then refracted back to the solar surface
within approximately 50 Mm of the source and within an hour of the
beginning of the flare.  The surface manifestation of this phenomenon is the appearance 
of ``ripples'' on the solar surface, which we identify as sunquakes. It is interesting to note that the majority of flares do not generate sunquakes. Most large flares are seismically inactive, 
which suggests that the strong magnetic fields of the hosting active
regions may substantially alter the behaviour of
helioseismic signals emerging from below. To date, the magnetism of solar
seismic regions has not been studied in depth. \citet{bdl1987} and
\citet{b1995} observed that sunspots partially absorb wave
energy and shift the phase of the oscillations. A long line of
theoretical developments then followed \citep{cb1993,cbz1994,bc1997,cc2003,cc2005}
which has shown that near-surface conversion to slow magnetoacoustic waves
is predominantly responsible for the absorption.

Furthermore, \citet{schunker2005}  confirmed through observations that 
magnetic forces should be of particular significance for acoustic signatures in penumbral regions,
where the magnetic field is significantly inclined from vertical.
\citet{sh2005} also found that a sizable proportion of magnetic field variations 
occur in the penumbral regions of flaring sunspots. Remarkably, the
majority of seismic sources induced by flares are also located either inside, or within close 
proximity to the penumbra. These observations possibly suggest a new mechanism which 
may be driving seismic waves at the photospheric level.  Indeed, \citet{hfw2007} have
recently introduced the idea of  the coupling of flare energy into a
seismic wave, namely the ``McClymont magnetic jerk'', produced during
the impulsive phase of acoustically active flares. They estimated the
mechanical work that would be done on the photosphere by a sudden
coronal restructuring. Their energy estimates are similar to those
based on our helioseismic observations.

During January 11\,--\,20 2005, AR10720 produced 5 X-class solar flares,
including an X7.1 on January 20 which produced an intense solar proton
storm. However, the Michelson Doppler Imager (MDI) onboard the Solar and Heliospheric Observatory (SOHO) instrument provided helioseismic observations only
for the X1.2 flare of January 15. This flare was situated at N14E08 on
the solar surface. The detection of the
powerful seismic transient of 2005 January 15 was initially reported by
\citet{betal2006} and \citet{Moradi2006a,Moradi2006b}. The properties of the seismic waves generated
by the event were later analysed by \citet{kz2006}.  \citet{moradi07}
(Paper I) extensively analysed the sunquake of 2005 January 15 and compared the acoustic signatures with other supporting
observations.  They also compared certain qualities exhibited by the
flare with all other known acoustically active flares. The coincidence between strong compact acoustic source and nearby signatures of hard X-ray emission is remarkable.  This and the spatial coincidence
of the acoustic emission with the sudden white-light signature,
suggests that the sudden heating of the low photosphere results in
 seismic waves at the solar surface.  \citet{moradi07} further 
suggested that a detailed examination of the heated magnetic
photosphere is needed to complete our studies. 

In this paper we will analyse the magneto-seismic activity of AR10720. Specifically, we 
shall investigate the role of the photospheric and coronal magnetic fields in generating the seismic
waves based on the results of \citet{hfw2007}.  We will also use vector
magnetograms of AR10720 to analyse the evolution and dynamics of the 
photospheric magnetic field.

The structure of this paper is as follows. Section 2 outlines the
observational data used for our analyses. Section 3 presents the
location of the seismic source in AR10720 with references to
\citet{moradi07} for details. Section 4 outlines
the line-of-sight magnetic transients in AR10720 associated with the
X1.2 solar flare.  Section 5 shows the coronal magnetic field
reconstruction models of AR10720.  In the final section, we conclude
with a discussion of the magnetism of the seismic region based on what
we have learnt from our analyses.

\section{Data}

The SOHO-MDI data consists of full-disk magnetograms obtained
at a cadence of 1 minute. The MDI data sets are described in more
detail by \citet{setal1995}. We analysed a dataset with a period of 2
hours encompassing the flare. We remapped the MDI images onto a
perspective that tracks solar rotation, with the region of interest
fixed at the centre of the frame. The MDI images were then Postel-projected
onto the frame with a nominal separation of 0.002 solar radii (1.4
Mm). The field of view in the MDI images analysed was
$256\times256$~pixels, thus incorporating a region of
$500\times500$~Mm on the solar surface.

MDI-magnetogram data provides a study of the structure and
variations of {\it l-o-s} (line of sight) magnetic fields in active
regions. Additionally, we have utilised photospheric vector magnetograms taken
by the Imaging Vector Magnetograph (IVM) instrument at Mees Solar Observatory, Hawaii \citep{mickey1996}.
These magnetograms provide the orientation and strength of the surface magnetic field in
AR10720.  The three magnetic components are: $B_{los}$ the
line-of-sight magnetic field component, $B_T$ and $B_{Az}$ the two transverse
components in the plane perpendicular to the line-of-sight. The
equations allowing the inversion of Stokes parameters introduce a
180$^o$ ambiguity on the azimuthal component $B_{Az}$ which can be
resolved using the method described in \citet{cetal1993}.

The Imaging Vector Magnetograph (IVM) provides the vector magnetic
field of AR10720 on January 14, 2005 at 18:08:12 UT, six hours before
the occurrence of the X1.2 flare.  Rotation and scaling were applied to
align the IVM data to the line-of-sight MDI magnetogram when
identifying the location of the seismic source on the IVM maps.

\section{Location of the seismic source}

\begin{figure}
\begin{center}
\includegraphics[width=0.95\columnwidth]{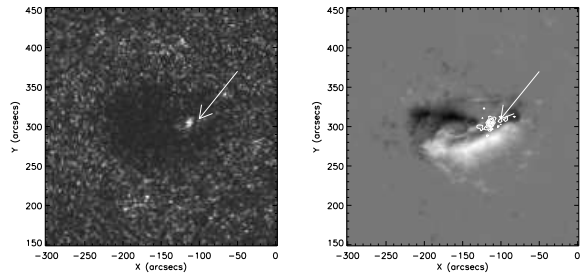}
\caption{Left: Snapshot of the egression power of AR10720 at 5\,--\,7~mHz on January 15 taken at the maximum of the seismic emission (00:44~UT). Right: SOHO-MDI magnetogram of the active region at 00:42~UT co-aligned with the acoustic source. The contour lines represent the overlaid acoustic source at 20, 40, 60, 80, 90 percent of the maximum intensity. The arrows indicate the location of the acoustic source.}

\label{fig1}
\end{center}
\end{figure}

In this section we briefly describe the main characteristics of the seismic
event generated by the flare of January 15, 2005. Paper I analysed
the general properties of the seismic source and identified some of the (possible) 
triggering mechanisms of this sunquake.

In Paper I, computational seismic holography was applied to MDI dopplergram
observations to image the seismic source of the flare. The resulting ``egression power maps" (Figure~\ref{fig1}, left frame) showed a
relatively compact seismic source surrounded by some diffuse emission.
The source was clearly visible in 2.5\,--\,4.5 mHz holographic images
and even more pronounced in 5\,--\,7 mHz images.  The conspicuous 6~mHz
seismic source, indicated by the arrows in Figure~\ref{fig1}, becomes
apparent near the western end of the active region at 00:33~UT,
reaching a maximum at 00:41~UT and disappearing at 00:47~UT.  The
source reveals two components: a compact kernel $\sim$10~Mm in
diameter on the magnetic neutral line and a diffuse spread, parallel to
the neutral line lenticular component, $\sim$45~Mm long (Paper~I).
These signatures correspond closely with other compact manifestations
of the flare (white light emission and magnetic kernels, see next section).
The suppression of ambient acoustic emission from the magnetic region
considerably enhances the significance of the acoustic emission from
the flare.

The powerful seismic waves produced by the sunquake had amplitudes exceeding 100~m/s, propagating 
with an elliptical shape with a major axis SE\,--\,NW \citep{kz2006}. The total energy
emitted by the 5\,--\,7 mHz seismic source was estimated at
$10^{27}$ergs. This is about the same as the seismic energy produced
by the October 28 (X17.2) flare and $\approx$ 200\% greater than the
October 29 (X10) flares \citep{dl2005}. Indeed, the 2005 January 15
flare contributes to recent findings that relatively small flares can
emit disproportionate amounts of acoustic energy \citep{dl2005,
moradi07}.

\begin{figure}
\begin{center}
\hspace*{-1cm}\includegraphics[width=1.25\columnwidth]{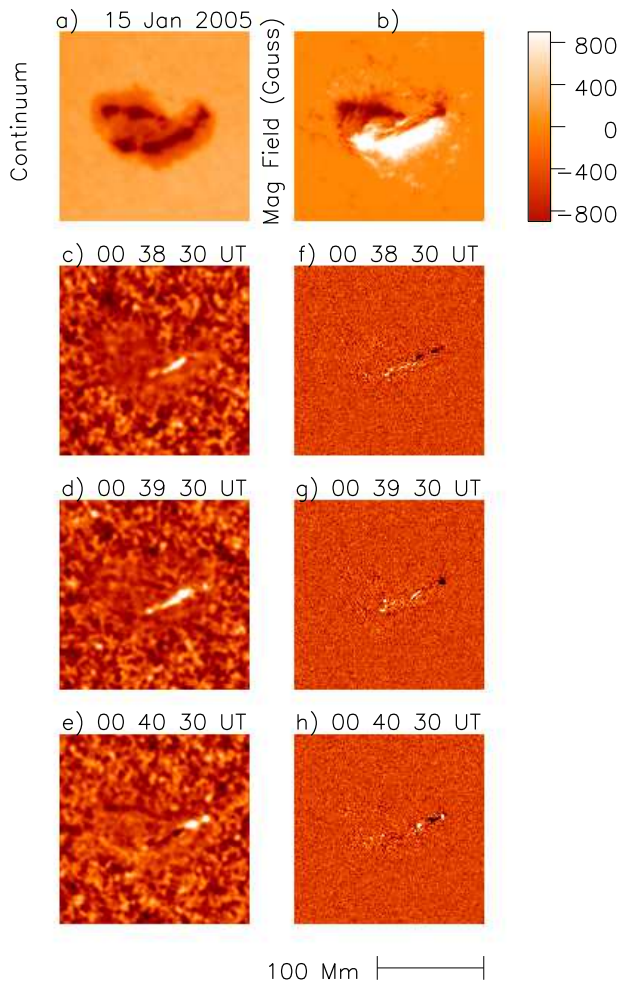}
\caption{GONG+ intensity continuum (panel a) and MDI magnetogram (panel b) images of AR10720 on 2005 January 15 at 00:39 UT. Panels c--e: differences between GONG+ intensity continuum images at the specified times. Panels f--h: differences between MDI magnetogram images.}
\label{fig2}
\end{center}
\end{figure}

In Figure~\ref{fig2} we show that the intensity continuum and magnetic
signatures of this flare spatially coincide.  The upper panels show a
GONG+ continuum image and a MDI line-of-sight magnetogram of AR10720
on January 15 at 00:39 UT.  Panels c, d and e show differences between
consecutive GONG+ intensity continuum images. For example, panel c
shows the differences of GONG+ images taken at 00:38 and 00:39~UT. The
subsequent two frames show consecutive differences 1 and 2 minutes
later. The right column (panels f, g and h) show also differences
between consecutive MDI magnetograms.  The visible continuum emission
is elongated along the magnetic neutral line, corresponding closely with
the lenticular component of seismic emission seen at 00:42~UT in
Figure~\ref{fig1} (left frame).  The magnetic kernels coincides
with both the seismic compact source, and the lenticular component.
\citet{moradi07} also reported that the continuum radiation into the seismic
area was $2 \times 10^{30}$~ergs, which is $\sim 500$ times the total
seismic energy we estimate the flare to have emitted into the
photosphere.

\section{Local Magnetic Fields in the Seismic Region}

\begin{figure}
\begin{center}
\includegraphics[width=0.9\columnwidth,height=0.6\columnwidth]{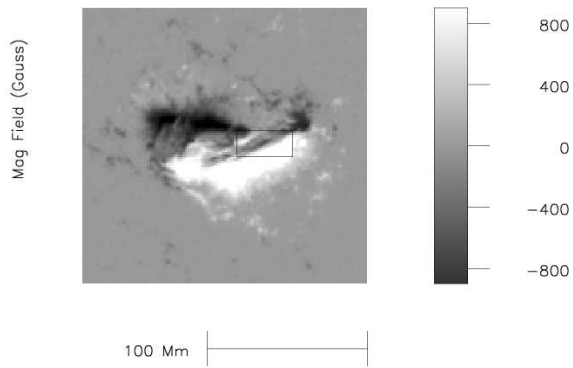}
\caption{MDI {\it l-o-s} magnetic field (in G) at 00:39 UT.  The rectangular region represent the highly seismic region of AR10720 (seismic area)}
\label{fig3}
\end{center}
\end{figure}

\begin{figure}
\begin{center}
\includegraphics[width=1.0\columnwidth, angle=0]{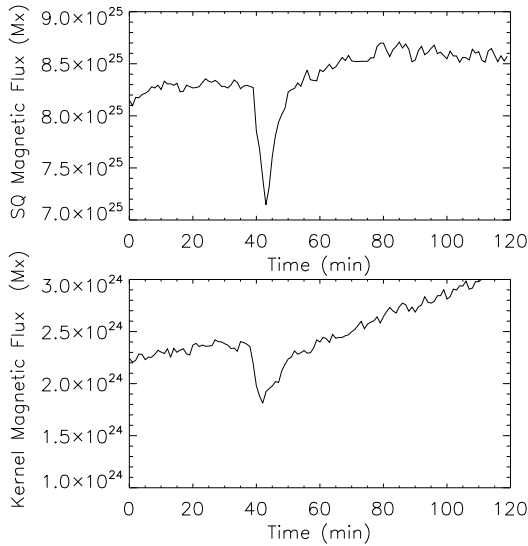}
\caption{ Magnetic flux during the flare integrated over:  Upper frame : the entire quaked (SQ)  region,  Lower frame: the acoustic main kernel. Minute 0 corresponds to 00:00 UT. }
\label{fig4}
\end{center}
\end{figure}

Once the position of the seismic sources are found, we 
 study  the structure and variation of the magnetic field at these locations.
This work may provide important 
information regarding the photospheric effects from solar flares. 
First, we analysed
{\it l-o-s} MDI magnetogram images taken during the seismic event. The
seismic source is identified in the vicinity of neutral lines
separating regions with opposite polarities of {\it l-o-s} magnetic
component (box in Figure~\ref{fig3}). Then, we extrapolated the
magnetic field lines based on the photospheric magnetograms.

Figure~\ref{fig4} shows the variation of the photospheric magnetic
field flux in the region of the seismic source (main kernel).  A sharp
decrease of the magnetic flux is observed during the impulsive phase
of the flare at 00:39 UT, followed by a gradual increase for about 10
minutes, before returning to the pre-flare magnetic flux levels. For the seismic
region, when averaging $B_{los}$, we used a rectangular area of $\sim
539$~Mm$^2$. These are reversible magnetic changes, similar to the
ones discussed by \cite{kz2001}.

The magnetic transients visualised in Figure~\ref{fig2} (right column)
are produced by precipitation of high energy particle beams  that
induce their own magnetic field and also change the thermal structure
of the photosphere, an effect that has an impact on the formation of the
NiI~$\lambda$6768~\AA~line (from which the $B_{los}$ is measured).

For the seismically active area of AR10720 we have detected
a number of $\mathrm 1.4 \times 1.4$~Mm seismic areas with abrupt and permanent changes in the magnetic field region as reported by
the statistics of \citet{sh2005} who proposed that these were
the result of significant changes in the longitudinal component of the
magnetic field.  We have also measured transient magnetic shifts
as seen in Figure~\ref{fig2}. These have also been detected in a
number of flares, some of which were acoustically active \citep{kz2001}.  The
magnetic signatures are spatially and temporally consistent with the
acoustic signature.

    The seismic emission occurred in a region where the magnetic field
is quite strong (field strengths in the range 400\,--\,1200~G) and
where the field lines are highly inclined (a range of 60-80 degrees)
to the vertical. For helioseismic purposes a strong magnetic field is
certain to be important throughout the photosphere and chromosphere,
particularly in penumbral regions, where the field is significantly
inclined from vertical \citep{schunker2005}.  Recent theoretical and
computational modelling of magnetised subphotospheres
\citep{cally2006} and atmospheres \citep{bogdan2002} has revealed that
fast-to-slow (or vice versa) magnetoacoustic wave conversion occurs
strongly near surfaces where the sound and Alfv\'en speeds coincide,
provided the local `attack angle' of the wave vector to the magnetic
field lines is fine. \citet{cg2008} have also found significant
conversion to the Alfv\'en wave. We will see in the next section that
indeed, the low-lying loops are not only highly inclined but strongly
twisted to facilitate the accumulation of the energy needed to trigger
the flare.

\section{Reconstruction of the 3-D magnetic field}

The acoustic activity of an active region is clearly related to
the structure of the coronal magnetic field, which facilitates the
precipitation of beams of non-thermal particles towards the
chromosphere. \citet{Martinez-Oliveros2007} suggested that the
coronal magnetic field configuration (height and symmetry of loops) can be a relevant
factor in the generation of photospheric seismic waves. They studied
the seismicity of the August 14, 2004 M7.4 solar flare, and found that
the seismic source was located just  beneath low-lying, highly-sheared magnetic
field lines.  This type of configuration seems to facilitate the
transport of flare energy into the photosphere.

\begin{figure}
\begin{center}
\includegraphics[width=0.85\columnwidth, angle=0]{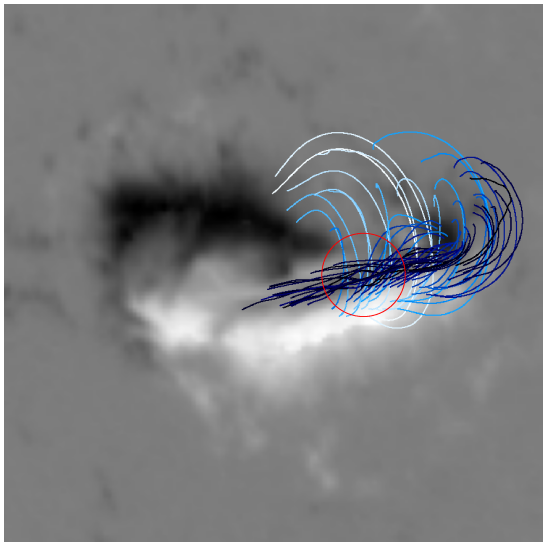}
\caption{A display of the NLFFFF  magnetic field extrapolation of AR10720 overlaid on the {\it l-o-s} IVM magnetogram. An intricate and complex structure of low-lying magnetic field lines are observed over the seismic region represented by the red circle.}
\label{3dima}
\end{center}
\end{figure}

In this work we have imaged the coronal magnetic field using
potential \citep{sakurai1982} and non-linear force free field (NLFFF)
extrapolations of the photospheric magnetic field based on the
optimization method of \citet{wrs2000}. The NLFFF extrapolation of the seismic region, shows
that the lower corona and upper chromosphere are dominated by magnetic field lines of
middle and low altitude (Figure~\ref{3dima}) which are highly twisted, similar to the
loops of the August 14, 2004 flare.

Figure~\ref{3dima} shows a map of the magnetic field lines
extrapolated with footpoints located at or close to the seismic
region. The map shows an intricate network of low-lying magnetic field
lines parallel to the magnetic neutral lines. This structure is
recognizable in the Transition Region and Coronal Explorer (TRACE) image at 1600~\AA (Figure~\ref{imaeit}).
The extrapolation was computed using the IVM vector magnetogram of
January 14, taken at 18:08 UT, 6 hours prior to the onset of the
flare. The close match between the visible flaring loops in the
TRACE observations and the NLFFF field lines shows that the magnetic
geometry did not change drastically in the six hours prior to the flare.

The complex structure of field lines suggests that the flare
sequentially illuminated the magnetic loop footpoints in some erratic
order but always localised in the same small area. Perhaps this
configuration, along with the impulsive characteristics of the flare
provided the necessary conditions to drive this powerful sun
quake. Complex and erratic motion of the HXR footpoints at the
location of the seismic source has been reported before
\citep{Martinez-Oliveros2007c, hwm2006, fhhm2007}.

\begin{figure}
\begin{center}
\includegraphics[width=0.95\columnwidth, angle=0]{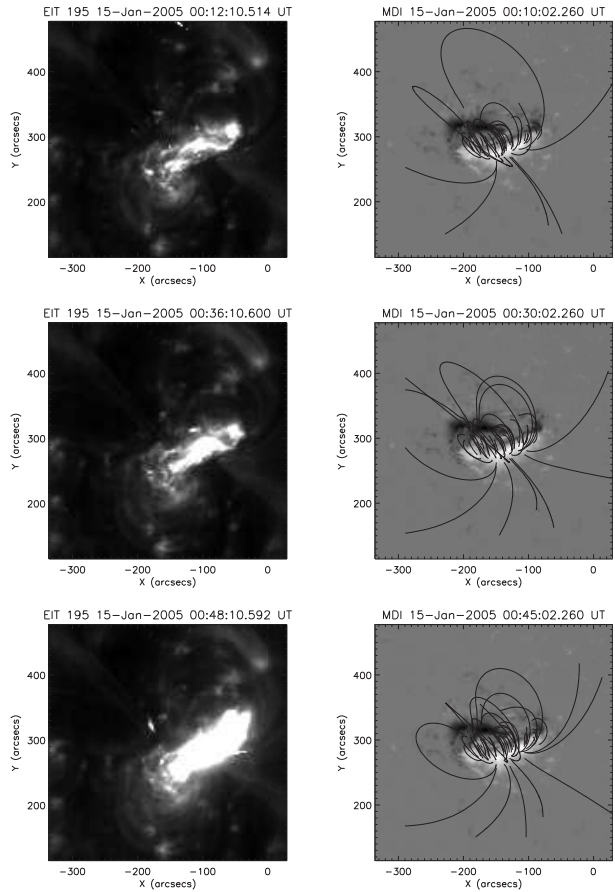}
\caption{Left column: Evolution of the coronal loops of  the active region 10720 seen in the SOHO-EIT images at 195~\AA. Right column: SOHO/MDI maagnetograms overlaid with the extrapolated potential magnetic field lines at the specified times.}
\label{potential}
\end{center}
\end{figure}

\begin{figure}
\begin{center}
\hspace*{-1cm}\includegraphics[width=1.35\columnwidth, angle=0]{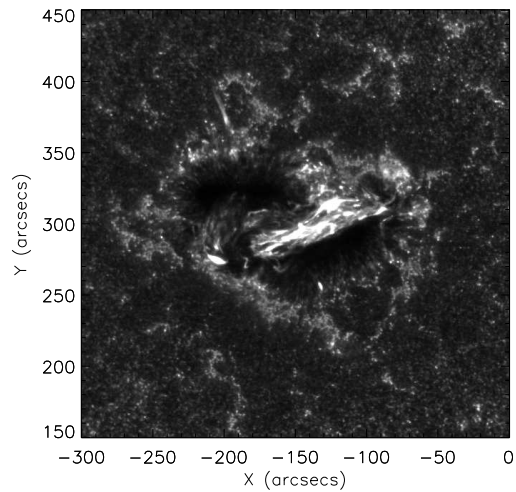}
\caption{TRACE image at 1600~\AA taken at 00:12:35~UT. The observable feature in the image resembles the structure observed in the extrapolation.}
\label{imaeit}
\end{center}
\end{figure}

Although, the NLFFF extrapolation gives a better approximation to the real configuration of the magnetic field, the magnetogram on which this method is based was obtained 6 hour prior to the flare. So, in order to obtain a more general description of the magnetic field configuration before, at the maximum, and after the flare, we calculate potential magnetic field extrapolations at different representative times, based on SOHO-MDI magnetograms (Figure~\ref{potential}). We focus our attention to the overall configuration of the magnetic field of the active region, comparing the results of the extrapolations with the observations made by Extreme ultraviolet Imaging Telescope (EIT) onboard SOHO at 195~\AA. We found that the extrapolations are dominated by high and medium altitude magnetic field lines, with mainly north--south orientation. This distribution of magnetic field lines are similar to those observed by SOHO-EIT. In figure~\ref{potential} we appreciate a redistribution of the magnetic field lines, that can be attribute to the reorganization of the photospheric magnetic field.

\section{Discussion}

We have shown that the seismic area in AR10720 is located just
beneath a complex coronal loop structure with highly twisted
lines, which means that  the photospheric impact was significant in
the region where the twisting allowed a maximum storage of energy. This
twisting  caused the interaction between neighbouring low-lying loops, triggering the flare. This can be seen in the SOHO-EIT images (Figure~\ref{potential}), where the
flaring of the loops is visible along the field lines. 

 A main question we want to ask now is: how was this sunquake produced? That is, what was the mechanism for transporting the flare energy efficiently, from the reconnection site downward into the
chromosphere?

Let us describe a number of possible mechanisms that could trigger a seismic source
at the photospheric level during a flare and discuss the likehood that these
mechanisms can explain the sunquake of the 2005 January 15 flare.

The first mechanism \citep{kz1998} propose that seismic emission
into the solar interior in sunquakes is the continuation of a
chromospheric shock and condensation resulting from explosive ablation
of the chromosphere and propagating downward through the photosphere
into the underlying solar interior. Chromospheric shocks are well
known under such circumstances, based on red-shifted H$\alpha$
emission at the flare site at the onset of the flare.  The simulations
were worked out at length by \citet{fcm1985a,fcm1985b,fcm1985c}
and others since.  The hypothesis that the photospheric emission is a
direct continuation of such shocks was considered by \citet{dl2005} and \citet{kz2006}. For the flare studied in this paper, the hydrodynamic impact of the photosphere was clearly
significant since, amazingly, this X1.2 type flare triggered a very
powerful seismic source and visible seismic waves (see Paper I and
\citet{kz2006} for details). The spatial coincidence between the
HXR emission and the seismic source leads us to connect the two
processes, and conclude that the high-energy electrons played an
important role.  However, we have to look at the statistics of
acoustically active events \citep{betal2008} and acknowledge that
most solar flares do not produce sunquakes. This leads us to
believe that, for the majority of flares, strong radiative damping
depletes the chromospheric transient before its arrival at the low
photosphere.  Therefore, we need to look for alternative mechanisms to explain the 
excitation of seismic sources.

In Paper~I we proposed a second mechanism. We stated that the coincidence between the
locations of sudden white-light emission and seismic emission in all acoustically active flares (including the 2005 January 15) suggests that a substantial component of the seismic emission
seen is a result of sudden heating of the low photosphere associated
with the observed excess of visible continuum emission (radiative back-warming). The
origin of white-light emission would have to be entirely in the
chromosphere, where energetic electrons dissipate their energy
\citep{metcalf1990,zk1991,zk1993}, mainly by
ionizing previously neutral chromospheric hydrogen approximately to
the depth of the temperature minimum. It appears that the low
photosphere itself would be significantly heated as well. This
is primarily the result of Balmer and Paschen continuum edge
recombination radiation from the overlying ionised chromospheric
medium, approximately half of which we assume radiates downward and
into the underlying photosphere.  \citet{dl2005}, \citet{donea2006}, and \citet{moradi07} have  analysed
this process in detail. \citet{cd2006} also affirm that the white-light
flare signatures highlight the importance of radiative back-warming in
transporting the energy to the low photosphere when direct heating by
beam electrons is impossible.

A third possible mechanism proposed by \citet{zharkova2007}, states that high-energy protons, can directly deposit energy in the photosphere, inducing a seismic source . However, for the flare of 2005 January 15 there is no indication of
high-energy protons that could directly supply the energy on which the
acoustic emission depends. Likewise, energetic electrons consistent
with HXR signatures, seem to be unable to penetrate into the low photosphere in
anywhere near sufficient numbers required to account for the direct heating needed
by the seismic sources \citep{mcs1990}.

A fourth mechanism \citep{hfw2007} suggested that the ``McClymont magnetic
jerk'' can account for the seismic activity of some flares.  Here, we
want to apply the relations of \citet{hfw2007} for the seismic area of
AR10720 in order to determine whether the ``McClymont magnetic jerk'' can
account for the seismic activity of the 2005 January 15 flare.

For a line-of-sight MDI magnetic field change of 60~G, as measured in
the region where the main kernel of the acoustic seismic appeared
(area $\sim 6$ Mm $ \times \, 9 $~Mm), the total Lorentz force, for
$B_z \sim 400$~G, is $2 \times 10^{21}$~dyne ($\delta f_z \sim 2.4
\times 10^3$~dyne/cm$^2$ $\cdot$ $1.2 \times 10^{18}$~cm$^2$). In
Paper I we observed that the photospheric impact produced a depression
of about 10 km. Using this, the maximum work done by the Lorentz force
on the photosphere is estimated at $\sim 2 \times 10^{27}$~ergs, which
is twice the energy needed by the entire seismic source to oscillate
at a frequency centered at 6 mHz within a 2 mHz band.  From Paper I we
extract that the seismic kernel accounted for $\sim 45$\% of the total
egression power (estimated at $\sim 1 \times 10^{27}$~ergs) integrated
over the region encompassing the entire flare signature (kernel plus
the lenticular diffuse component).  Of course, the inferred number is
just an upper limit based on many uncertainties of the local
physics. We conclude that the ``McClymont magnetic jerk'' may explain
the formation of the acoustic kernel, but does not explain the diffuse
lenticular element of seismic activity surrounding the main kernel,
which is distributed along the neutral line up to $\sim$15~Mm east and
$\sim$30~Mm west of the kernel \citep{moradi07}. The fact that the
erratic motion of the HXR sources is observed only above the acoustic
kernelled area sustains this assumption.

 We  note that if integrating over the full area of the seismic
source (including the diffuse lenticular acoustic emission surrounding
the main kernel) the change in magnetic field is very small (about 5~G),
which is understandable, because the full area of the seismic source
spans negative and positive magnetic polarities.  The area is also
permeated by field lines from loops of different orientation, making
the local magnetic geometry much more complicated. We also emphasize
that the seismic area of the solar flare of 2005 January 15 has
magnetic loops of a very large inclination angle, positioning the
reconnection site close to photospheric levels. 

According to \citet{h2000} and \citet{hfw2007}, one expects that the
field in the photosphere should become ``more horizontal'', as a
result of the coronal magnetic field contraction that follows the
decrease in the coronal magnetic energy.  Limited by the existing
observations, we cannot say whether the overall field structure of
AR10720 had tilted even more during the flaring.  Clearly, we cannot
definitively affirm that for this complicated structure of AR10720 the
Lorentz force is the main triggering mechanism for this quake (we
remind the reader that the X1.2 flare of 2005 January 15 generated the
most powerful solar seismic source detected so far). We believe that in
reality, a combination of all the above mechanisms may be required to
describe the entire phenomenon.

For simpler magnetic field configurations, where seismic sources have
 been also identified \citep{dl2005, donea2006} as localised acoustic
 kernels at the location of moving hard X-ray footpoints, we expect
 the ``McClymont jerk'' mechanism to work efficiently in parallel with
 the chromospheric shocks driven by sudden, thick-target heating of
 the upper and middle chromosphere \citep{kz1998,dl2005} and the
 ``back-warming'' mechanism.

\label{lastpage}

\end{document}